\documentclass[english,pra, twocolumn]{revtex4-1}
\usepackage[T1]{fontenc}
\usepackage[latin9]{inputenc}
\usepackage{textcomp}
\usepackage{amsmath}
\usepackage{amssymb}
\usepackage{graphicx}

\makeatletter
\usepackage[colorlinks = true,
            linkcolor =red,
            urlcolor  =magenta,
            citecolor = blue,
            anchorcolor = blue]{hyperref}

\makeatother

\usepackage{babel}
\begin{document}
\title{Nonclassicalities of hybrid coherent states }
\author{Yusuf Turek$^{1}$}
\email{yusuftu1984@hotmail.com}

\author{Nuerbiya Aishan$^{2}$ }
\author{Akbar Islam$^{3}$}
\affiliation{$^{1}$School of Physics Science and Technology, Xinjiang University,
Urumqi, Xinjiang 830046, China}
\affiliation{$^{2}$Hami Branch of China Mobile Group Xinjiang Co.Ltd, Hami, Xinjiang,
839000, China}
\affiliation{$^{3}$School of Physics and Electronic Engineering, Xinjiang Normal
University, Urumqi, Xinjiang 830054, China}
\date{\today}
\begin{abstract}
We address nonclassicality of hybrid coherent states (HCS), i.e. states
expressed as superpositions of coherent states and single-photon-added
coherent (SPAC) state. In particular, we evaluate their photon statistics,
squeezing, and negativity of the Wigner function. Our results indicated
that HCS may exhibit larger nonclassicalities than SPAC state. We
also suggest a generation scheme for HCS which involves Kerr nonlinearity
and postselection.
\end{abstract}
\maketitle

\section{\label{sec:1}Introduction}

States of the radiation field showing nonclassical behaviour play
a relevant role in the development of quantum information and quantum
technology. Among them, the generation of non-Gaussian states is an
important and challenging task in quantum state engineering. Over
the last decades, various schemes for the generation and optimization
of various quantum states including Fock states \citep{PhysRevLett.70.762,PhysRevLett.88.143601,RN1929},
Schrödinger cat states \citep{PhysRevA.55.3184,PhysRevA.72.022320,75,RN1933,2007Generation,74,PhysRevLett.121.143602,RN1935,RN1936,RN1937,RN1934},
squeezed states \citep{RN1932}, photon number states \citep{PhysRevLett.56.58,PhysRevA.36.4547,PhysRevA.39.3414,RN1930,Liu_2004,Waks2006,RN1931},
binomial states \citep{79,RN1940,78,76}, and squeezed state excitations
\citep{PhysRevLett.97.083604,PhysRevLett.101.233605,Liu2015,RN1938}
have been proposed and implemented. Furthermore, it has been shown
that the addition or subtraction of photons can change the physical
properties of a given state \citep{2007Physics}, and effective methods
for generating photon-added or subtracted states have been realized
using conditional measurements on a beam splitter\citep{1998Quantum,Mattos:22,PhysRevA.104.033715}.
The photon-added coherent states \citep{PhysRevA.43.492,PhysRevA.91.022334,PhysRevA.83.035802},
photon-subtracted or -added squeezed states \citep{PhysRevA.75.032104,PhysRevA.91.022317,PhysRevA.82.043842}
have attracted interest for their potential applications in several
protocols for quantum information processing\citep{C12}. Those states
can be produced by repeated applications of photon creation or annihilation
operators \citep{PhysRevA.72.033822}, respectively, on a given states
\citep{142,PhysRevA.72.023820,PhysRevA.74.033813,PhysRevA.82.063833,xu2019}.
Among them, photon-added coherent states \citep{PhysRevA.43.492}
are of particular interest since the process of photon addition take
states that are initially semiclassical and Gaussian, and lead to
non-Gaussian states with nonclassical properties including sub-Poissonian
photon statistics, squeezing and negativity of Wigner function.

The single-photon-added coherent (SPAC) state $\frac{a^{\dagger}\vert\alpha\rangle}{\sqrt{1+\vert\alpha\vert^{2}}}$
is generated by adding one photon to coherent state \citep{PhysRevA.82.063833}.
It reduce to single photon state for $\vert\alpha\vert$\textrightarrow 0
and to coherent states for $\vert\alpha\vert$\textrightarrow $\infty$
\citep{142,PhysRevA.72.023820}. Given that SPAC state have no vacuum
component and contain large single-photon probability, it find applications
in many fields as quantum state engineering \citep{129}, quantum
communication \citep{RN1923}, quantum key distribution \citep{RN1927,PhysRevA.90.062315,WANG20171393,RN1920},
and quantum metrology \citep{130,gard2016photon,PhysRevA.90.013821,SCHNABEL20171}.
Furthermore, recently \citep{RN1921} it was noticed that the performance
of quantum digital signature protocols can be improved significantly
using SPAC states instead of weak coherent states. However, when $\vert\alpha\vert$
is large, the SPAC states reduce to coherent states, and their non-Gaussianity
is lost, and thus SPAC states cannot be used directly in applications.
Besides, obtaining the non-linearity required to generate nonclassic
and non-Gaussian states is often challenging. At the same time, noise
and other imperfections in the optical setup may hinder the effective
generation of nonclassicality. To overcome these challenges, generating
nonclassical states, e.g., SPAC states, is usually obtained by conditionally
exploiting postselection \citep{PhysRevA.43.492,142}. In these schemes,
the initial coherent state component appears naturally at the output
in case of "negative postselection", and one may wonder whether
the superposition of SPAC and coherent states may be obtained and
exploited \citep{PhysRevA.82.053812,PhysRevA.91.022334,PhysRevA.91.053820,PhysRevA.98.011801,PhysRevA.100.043802,PhysRevA.102.043709,PhysRevA.104.033715}. 

The mathematical expression of this kind of states, which will be
referred to as HCS is 
\begin{equation}
\vert\psi\rangle=\mathcal{N}\left[\sqrt{\varepsilon}e^{i\theta}\vert\alpha\rangle+\sqrt{1-\varepsilon}e^{i\varphi}a^{\dagger}\vert\alpha\rangle\right]\label{eq:1}
\end{equation}
where the normalization constant $\mathcal{N}$ is given by 
\begin{equation}
\mathcal{N}=\left[2\sqrt{\varepsilon-\varepsilon^{2}}Re[e^{i(\theta-\varphi)}\alpha]+(1-\varepsilon)\vert\alpha\vert^{2}+1\right]^{-\frac{1}{2}}.\label{eq:2}
\end{equation}
Here, $\varepsilon$ is the real parameter bound to $0\le\varepsilon\le1$,
and $\vert\alpha\rangle$ is coherent state defined as 
\begin{equation}
\vert\alpha\rangle=D(\alpha)\vert0\rangle=\exp[\alpha a^{\dagger}-\alpha^{\ast}a]\vert0\rangle\label{eq:3}
\end{equation}
with $\alpha=re^{i\omega}$. From this expression, we can see that
the transition from SPAC to a coherent state can be accomplished by
adjusting the parameter $\varepsilon$ from zero to one. This kind
of states has been discussed in \citep{PhysRevA.100.043802}, where
a generation scheme involving an optical parametric amplifier has
been suggested. No photon addition is involved, and amplitude $\varepsilon$
can be controlled by varying the gain of the amplifier, see Eq.(7)
of\textit{ \citep{PhysRevA.100.043802}}{]}. In \citep{PhysRevA.105.022608}
a scheme to generate SPAC state via postselected weak measurements
in a three-wave mixing has been proposed, and the output state has
the HCS form of Eq. (\ref{eq:1}). The state $\vert\psi\rangle$ as
a coherent superposition between the SPAC state and coherent state
may possess some interesting features that could be used in some quantum
information processes. Furthermore, a recent work \citep{Mishra_2021}
revealed that the existence of an optical field vacuum in the could
change the non-classicalities of the Schrödinger cat states. However,
a question arises as to whether superpositions of SPAC and coherent
states also have more interesting nonclassical properties compared
to the SPAC states themselves.

This paper investigates the nonclassical properties of HCS and finds
that, indeed, they are characterized by larger non-classicalities
than SPAC state. Our figures of merit include the sub-Possionian character
of the photon statistics, the presence of squeezing, and the negativity
of the Wigner function. In particular, we investigate how for $\varepsilon=0.5$
and $\varepsilon=0.75$ strong non-classicality may also be observed
when $\vert\alpha\vert$ is larger than one (a "large" value in
this context). Our results may be used to assess and optimize generation
schemes for SPAC and HCS states in quantum information applications.

The remainder of this paper is organized as follows. Sec. \ref{sec:2}
analyzes in detail the photon statistics, squeezing, and Wigner function
of HCS and discuss the corresponding non-classicalities under different
conditions. Sec. \ref{sec:3} elaborates on a conditional scheme for
generating HCS, and finally, Sec. \ref{sec:4} concludes this paper.
It should be noted that throughout the paper, we use the unit $\hbar=1$.

\section{\label{sec:2}Nonclassical properties of hcs }

This section investigates the quantum statistical properties of HCS.
Specifically, we evaluate photon distribution, the Mandel Q-factor,
skew information, quadrature squeezing, and amplitude-squared (AS)
squeezing parameters and investigate the properties of the Wigner
function. 

\subsection{Photon statistics }

\begin{figure}
\includegraphics[width=8cm]{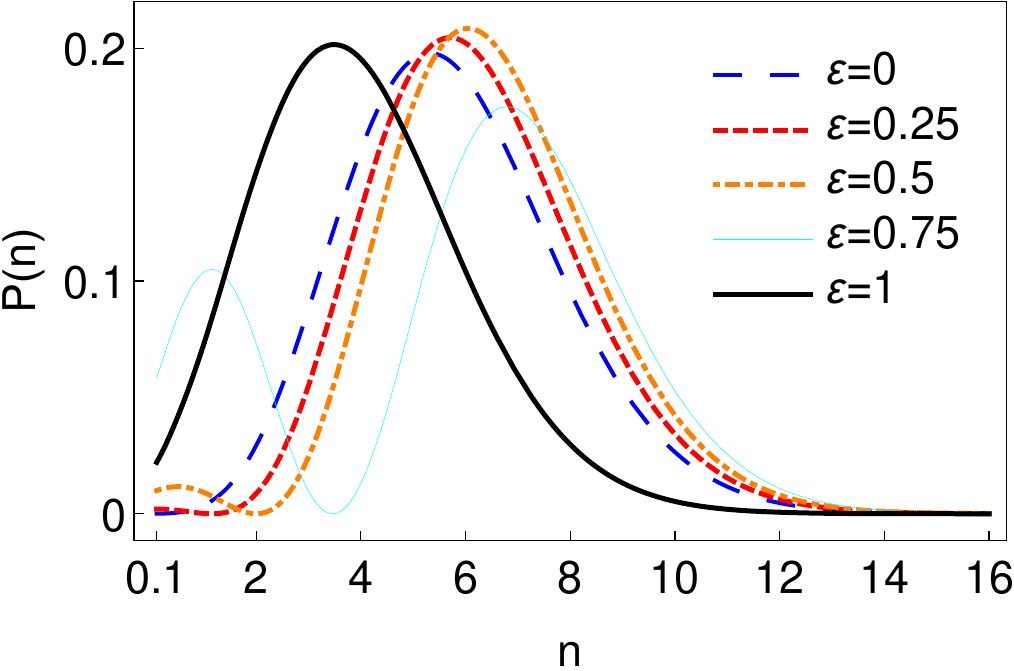}

\caption{\label{fig:1}Photon distribution of HCS states for different values
of $\varepsilon$. We have set $\omega=\varphi=0$, $\theta=\pi$,
and $\vert\alpha\vert=2$.}

\end{figure}

The photon distributions of a HCS is given by 
\begin{align}
P_{n} & =\vert\langle n\vert\psi\rangle\vert^{2}\nonumber \\
 & =\left|\mathcal{N}e^{-\frac{\vert\alpha\vert^{2}}{2}}\left(\sqrt{\varepsilon}e^{i\theta}\frac{\alpha^{n}}{\sqrt{n!}}+\sqrt{1-\varepsilon}e^{i\varphi}\sqrt{\frac{n}{(n-1)!}}\alpha^{n-1}\right)\right|^{2}.\label{eq:56}
\end{align}
To check the probability of finding $n$ photons in state $\vert\psi\rangle$,
we plot the $P_{n}$ as a function of photon number $n$ for different
values of $\varepsilon$ and the results are illustrated in Fig. \ref{fig:1}.
It is known that the photon distribution of a coherent state obeys
the Poisson distribution with a mean of $\vert\alpha\vert$, and for
large $\vert\alpha\vert$ the Poisson distribution can be approximated
as a Gaussian \citep{Agarwal2013}. The Fig. \ref{fig:1} illustrates
the distribution of Eq. (\ref{eq:56}) for $\vert\alpha\vert=2$ and
different values of $\varepsilon$. The dark solid line is obtained
for $\varepsilon=1$, i.e., the Poissonian distribution of a coherent
state with $\vert\alpha\vert=2$. The blue dashed curve corresponds
to $\varepsilon=0$, and represents the photon distribution of a SPAC
state. The other curves, corresponding to $\varepsilon\in(0,1)$,
illustrate the photon distributions of the HCS states.

For the unbalanced superposition with $\varepsilon=0.75$, (SPAC state
and coherent state dominates 25\% and 75\%, respectively), we observe
photon distribution oscillation of the corresponding states caused
by the interference effect of the two conditions. The width of the
distribution for $\varepsilon=0.75$ is narrower than for a coherent
state ($\varepsilon=1$ case) and suggests that it may be sub-Poissonian,
signaling the non-classicality of the corresponding HCS state.

\subsubsection{$Q_{m}$ Factor }

\begin{figure}
\includegraphics[width=8cm]{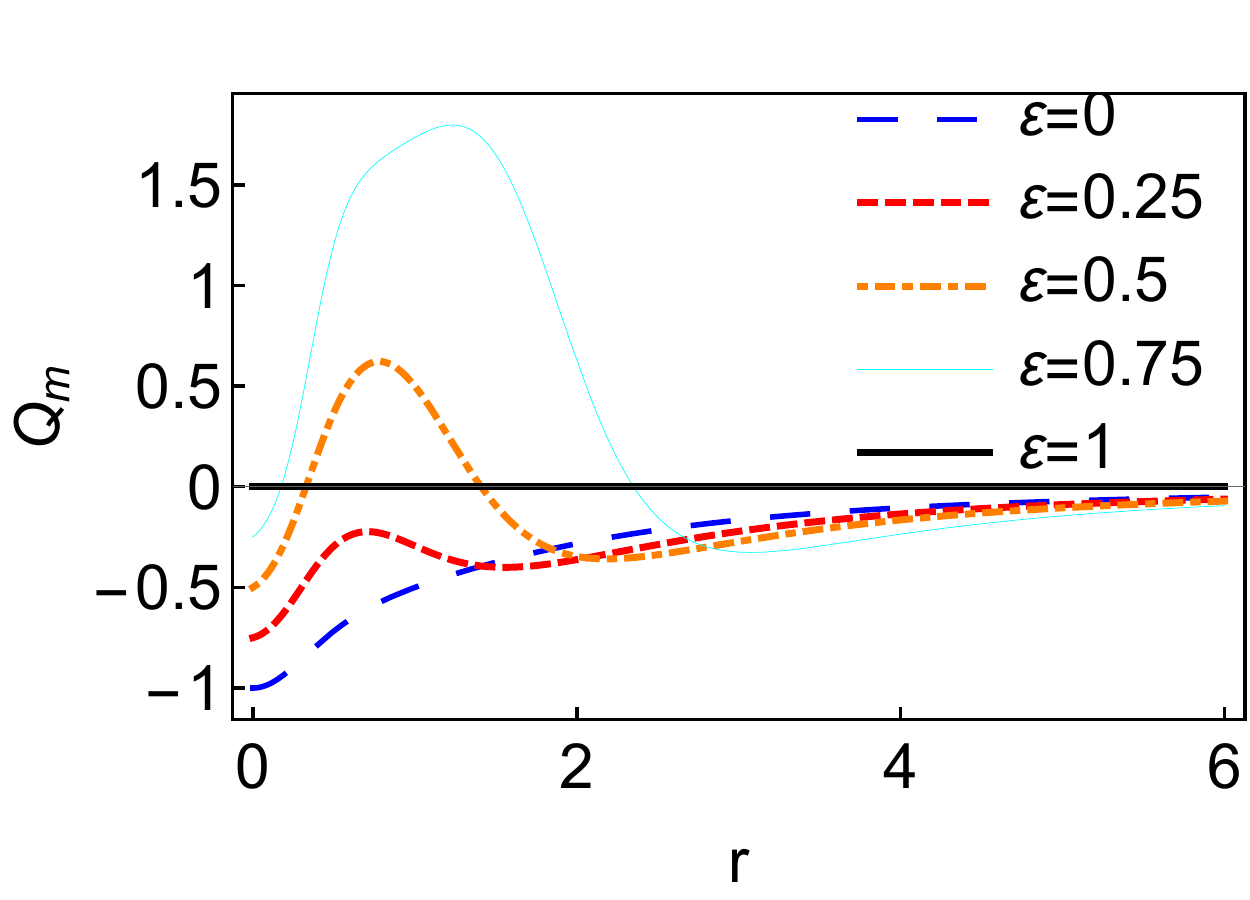}

\caption{\label{fig:2}Mandel factor of HCS as function of coherent state parameter
$r$ for different values of the superposition parameter $\varepsilon$.
The other parameters are set as in Fig.\ref{fig:1}. }

\end{figure}

The Mandel $Q$ factor, which can quantifies the sub-Poissonian character
of a distribution, is given by \citep{Agarwal2013} 
\begin{equation}
Q_{m}=\frac{\langle(\triangle n)^{2}\rangle-\langle\hat{n}\rangle}{\langle\hat{n}\rangle}=\frac{\langle a^{\dagger}a^{\dagger}aa\rangle-\langle a^{\dagger}a\rangle^{2}}{\langle a^{\dagger}a\rangle}.\label{eq:MF}
\end{equation}
 where $Q_{m}=0$ and $Q_{m}=-1$ are the values for a coherent and
a Fock state (maximally nonclassical state), respectively. The above
mathematical expression of Mandel's factor suggests that it can be
used as a natural measure of the variance departure of the photon
number from the Poisson distribution. As mentioned above, Mandel's
factor of coherent state equals zero, and it has Poissonian photon
statistics. The negativity of $Q_{m}$ is a sufficient condition for
the state to be non-classical since the corresponding photon number
variance is less than the average photon number, and the distribution
is sub-Poissonian. If $Q_{m}>0$, no conclusion can be made about
the non-classicality of the state. In order to investigate the $Q_{m}$
factor of HCS, we calculate the explicit expressions of $\langle a^{\dagger}a\rangle$
and $\langle a^{\dagger2}a^{2}\rangle$, i.e. 
\begin{align}
\langle a^{\dagger}a\rangle & =\mathcal{N}^{2}\left[(3-2\varepsilon)\vert\alpha\vert^{2}+(1-\varepsilon)\vert\alpha\vert^{4}-\varepsilon+1\right.\nonumber \\
 & \left.+2\sqrt{\varepsilon-\varepsilon^{2}}(1+\vert\alpha\vert^{2})Re[e^{i(\theta-\varphi)}\alpha]\right],\label{eq:AA}
\end{align}
and 
\begin{align}
\langle a^{\dagger2}a^{2}\rangle & =\mathcal{N}^{2}\left[(4-4\varepsilon)\vert\alpha\vert^{2}+(5-4\varepsilon)\vert\alpha\vert^{4}+(1-\varepsilon)\vert\alpha\vert^{6}\right.\nonumber \\
 & \left.+2\vert\alpha\vert^{2}\sqrt{\varepsilon-\varepsilon^{2}}(2+\vert\alpha\vert^{4})Re[e^{i(\theta-\varphi)}\alpha]\right]\,,\label{eq:7a}
\end{align}
and then input them in Eq. (\ref{eq:MF}), with the corresponding
results illustrated in Fig. \ref{fig:2}.

The SPAC states are non-classical and, in turn, their $Q_{m}$ factor
lies in the range $-1\leq Q_{m,spacs}<0$ as far as $\vert\alpha\vert${[}
is not too large (blue dashed curve in Fig. \ref{fig:2}){]}. Fig.
\ref{fig:2} highlights that while the $Q_{m}$ factor of the HCS
state is negative for many parameters. This might not always be valid.
Indeed, referring to Fig. \ref{fig:2}, we see that for $\varepsilon=0.5$
and $\varepsilon=0.75$, the $Q_{m}$ factor may be positive in some
regions. On the other hand, for one of the very same values ($\varepsilon=0.75$)
the corresponding HCS may take large negative values for larger $\vert\alpha\vert$.
Therefore, we conclude that tuning the superposition parameter $\varepsilon$
may provide large non-classicality, even larger than the corresponding
SPAC state, due to interference between the coherent component and
the photon-added one.

\subsubsection{Wigner-Yanase skew information}

\begin{figure}
\includegraphics[width=8cm]{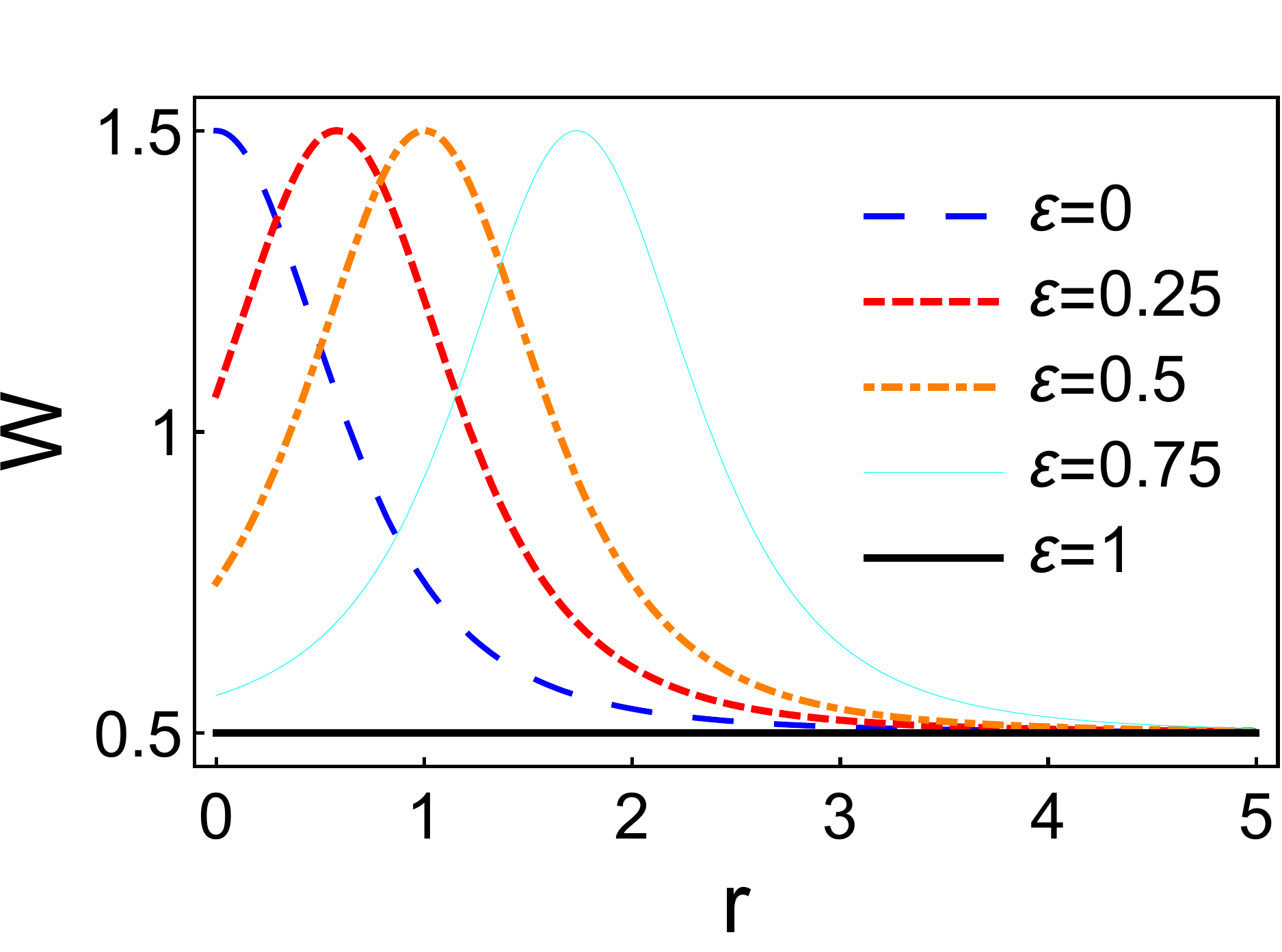}

\caption{\label{fig:W} Skew information relative to HCS as a function of coherent
state parameter $r$ for different values of the superposition parameter
$\varepsilon$. The other parameters are set as in Fig.\ref{fig:1}. }

\end{figure}

By virtue of Wigner-Yanase skew information, the recent studies\citep{1963,PhysRevA.100.032116}
introduced another information-theoretic quantifier for non-classical
light, which is conceptually simple, mathematically computable, physically
intuitive. For a pure, single mode state of the radiation field $\vert\psi\rangle,$the
skew information is written as \citep{PhysRevA.100.032116}
\begin{equation}
W\left(\vert\psi\rangle\right)=\frac{1}{2}+\langle\psi\vert a^{\dagger}a\vert\psi\rangle-\langle\psi\vert a^{\dagger}\vert\psi\rangle\langle\psi\vert a\vert\psi\rangle.\label{eq:8}
\end{equation}
 From this mathematical expression we can deduce that the skew information
is non-negative and for pure states has a minimum value of $W_{min}=\frac{1}{2}$
(coherent state). The larger values of $W$ indicate a larger nonclassical
character of a given state \citep{PhysRevA.100.032116}. We investigated
the skew information $W$ of the HCS by calculating the related quantities
in Eq. (\ref{eq:8}). In Fig. \ref{fig:W}, we have plotted the W
as a function of coherent state parameter $r$ for different values
of the superposition parameter $\varepsilon$ while setting $\omega=\varphi=0$
and $\theta=\pi$. As showed in Fig. \ref{fig:W}, the skew information
of SPAC state ($\varepsilon=0$) is decreased with increasing$\vert\alpha\vert$
and reached $0.5$ since the SPAC state is reduced to coherent state
as $\vert\alpha\vert$\textrightarrow $\infty$. Furthermore, another
very interesting phenomena we observed in Fig. \ref{fig:W} is that
the skew information of HCS for $\varepsilon\neq0,1$ cases have high
$W$ values in regions where the $W$ value of SPAC state decreased.
Thus, the Wigner-Yanase skew information of HCS captures the nonclassical
features of HCS, and it can be associated to sub-Poissonian statistics. 

\subsection{Squeezing }

Squeezing is another typical feature of non-classical states. To further
check the non-classical properties of HCS in Eq. (\ref{fig:1}), in
this subsection we study the occurrence of the "ordinary" quadrature
squeezing or of the amplitude-squared squeezing in our states. 

\subsubsection{Quadrature Squeezing }

\begin{figure}
\includegraphics[width=8cm]{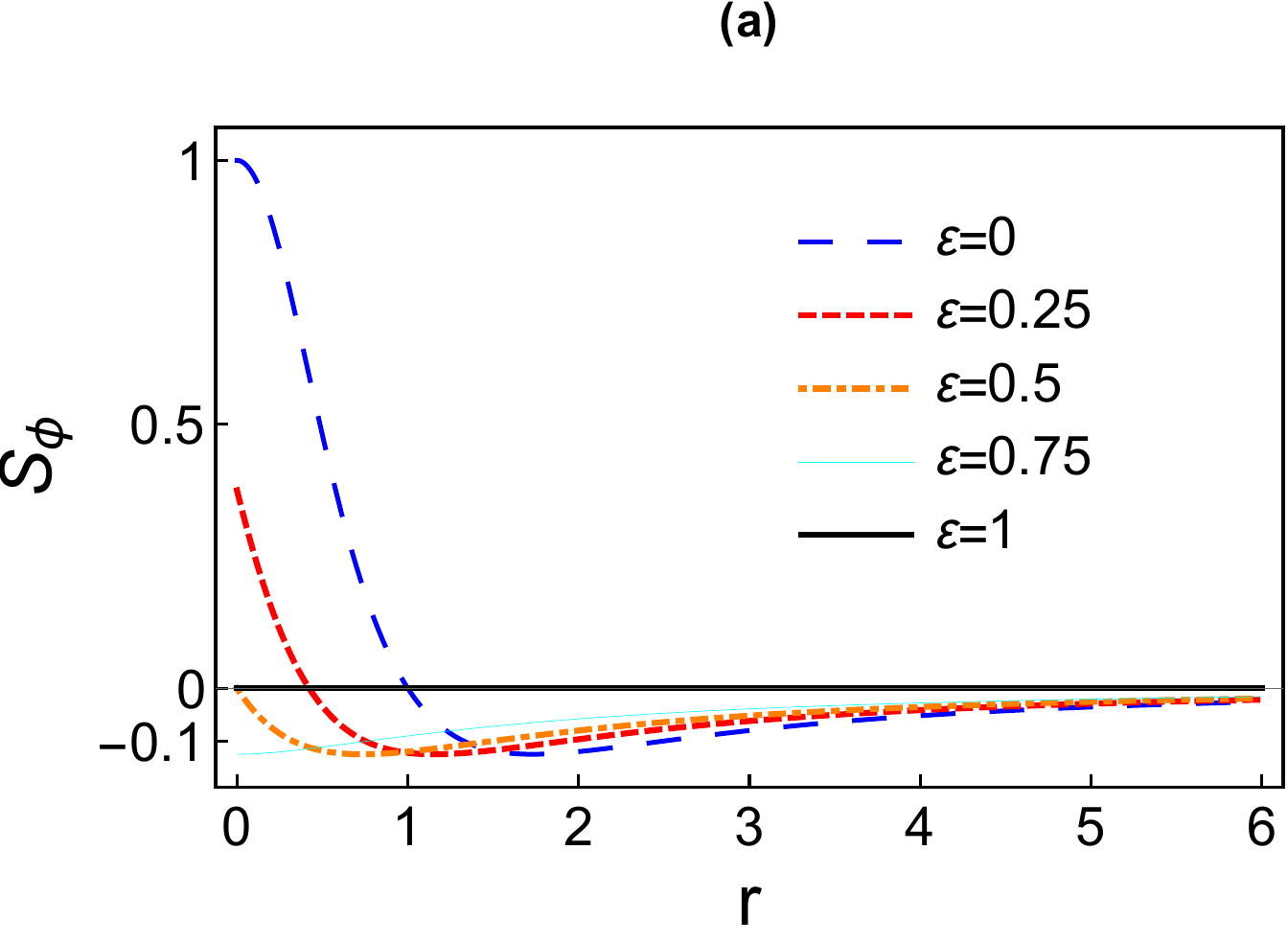}

\includegraphics[width=8cm]{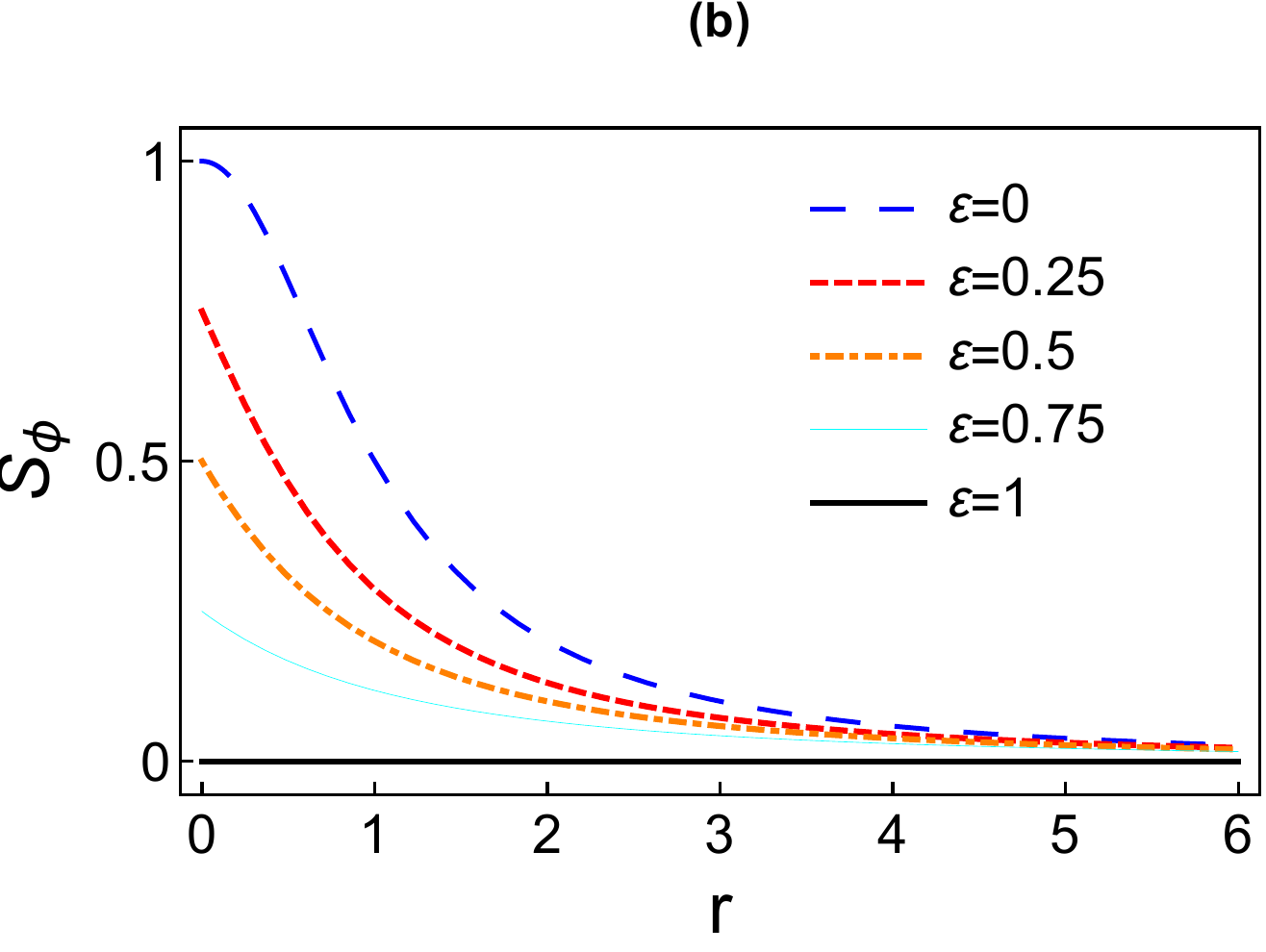}

\caption{\label{fig:3}Quadrature squeezing parameter of HCS as a function
of the modulus of the coherent amplituded $\vert\alpha\vert$, for
different values of the superposition parameter $\varepsilon$. We
set $\omega=\theta=\varphi=0$, and $\phi=0$ in (a), $\phi=\frac{\pi}{2}$
in (b).}

\end{figure}

Quadrature squeezing is a non-classical effect that may be quantified
by the single-mode squeezing parameter 
\begin{equation}
S_{\phi}=\langle:X_{\phi}^{2}:\rangle-\langle X_{\phi}\rangle^{2}\label{eq:15}
\end{equation}
where 
\begin{equation}
X_{\phi}=\frac{1}{\sqrt{2}}\left(ae^{-i\phi}+a^{\dagger}e^{i\phi}\right),\ \ \ [X_{\phi},X_{\phi+\frac{\pi}{2}}]=i\label{eq:16}
\end{equation}
denotes a quadrature operator of the field, and $:\ :$ stands for
normal ordering of operator defined by $:aa^{\dagger}:=a^{\dagger}a$,
whereas $aa^{\dagger}=a^{\dagger}a+1$. We observe that $S_{\phi}$
is related to the variance of $X_{\phi}$ as: 
\begin{equation}
S_{\phi}=\left(\triangle X_{\phi}\right)^{2}-\frac{1}{2}\label{eq:17}
\end{equation}
where $\triangle X_{\phi}=\sqrt{\langle X_{\phi}^{2}\rangle-\langle X_{\phi}\rangle^{2}}$.
For a coherent state, the squeezing parameter equals zero, whereas
the minimum value is $S_{\phi}=-\frac{1}{2}$, and for a nonclassical
state $S_{\phi}\in[-\frac{1}{2},0)$, since it is semi-classical state
satisfying the minimum uncertainty relation. The explicit expression
of $S_{\phi}$ may be easily obtained from Eqs. (\ref{eq:17}), and
its behavior as a function of $\vert\alpha\vert^{2}$ for different
values of $\varepsilon$ is presented in Fig. \ref{fig:3}, where
we set $\omega=\theta=\varphi=0$. The latter figure highlights that
the squeezing appears for the quadrature $X_{\phi=0}$. From Fig.
\ref{fig:3}(a), we see that squeezing occurs for all HCS states for
a large enough $\vert\alpha\vert$ and then decreases for increasing
$\vert\alpha\vert$, vanishing asymptotically. We also observe that
by increasing $\varepsilon$, a larger squeezing is obtained for smaller
$\vert\alpha\vert$, as compared to the $\varepsilon=0$ case. Thus,
the numerical results indicated that rather than the coherent ($\varepsilon=1$)
or SPAC state ($\varepsilon=0$) exist independently, their superposition
affords a very good squeezing effect. Especially, the state corresponding
to $\varepsilon=0.75$ has a better squeezing effect in the $X_{\phi=0}$
quadrature compared to the SPAC state, whereas form Fig. \ref{fig:3}(b),
we see that for the $X_{\phi=\pi}$ quadrature we have no squeezing
for all of the values of $\varepsilon$ and $\vert\alpha\vert$.

\subsubsection{Amplitude-squared(AS) squeezing}

Squeezing of squared field-amplitude (amplitude-squared squeezing)
has been shown to be a nonclassical effect and some specific examples
has been studied \citep{33,32,31}. In order to introduce AS squeezing,
let us consider the real and imaginary parts of the square of the
field mode amplitude, i.e. 
\begin{equation}
Y_{1}=\frac{\left(A^{\dagger2}+A^{2}\right)}{2},\ \ Y_{2}=i\frac{\left(A^{\dagger2}-A^{2}\right)}{2},\label{eq:Y12}
\end{equation}
where $A$ and $A^{\dagger}$ are slowly varying operators defined
by $A=e^{i\vartheta t}\hat{a}$, $A^{\dagger}=e^{-i\vartheta t}\hat{a}^{\dagger}$,
and obey canonical commutation relations. The operators $Y_{1}$ and
$Y_{2}$ satisfy the commutation relation \citep{32,2013}
\begin{equation}
\left[Y_{1},Y_{2}\right]=i\left(2N+1\right)\text{,}\label{eq:commutation relationship}
\end{equation}
where $N$ is the number operator, $N=A^{\dagger}A$. Thus, the operators
$Y_{1}$ and $Y_{2}$ obey the uncertainty relation 
\begin{equation}
\Delta Y_{1}\triangle Y_{2}\ge\langle N+\frac{1}{2}\rangle.\label{eq:uncertanty relation}
\end{equation}
Here, $\triangle Y_{1,2}=\sqrt{\langle Y_{1,2}^{2}\rangle-\langle Y_{1,2}\rangle^{2}}$
denotes the variance of $Y_{1,2}$ over a generic state $\vert\phi\rangle$.
We say that the AS squeezing exists in the variable $Y_{i}$ if it
satisfies 
\begin{equation}
\left(\text{\ensuremath{\triangle}}Y_{i}\right)^{2}\text{\ensuremath{<}}\langle N+\frac{1}{2}\rangle\text{\ \ \ \ for \ensuremath{i=1} or \ensuremath{2}.}\label{eq:squeezing}
\end{equation}
In general, the AS squeezing factor can be defined as 
\begin{equation}
Y=\left\langle \left(Y_{\vartheta}-\langle Y_{\vartheta}\rangle\right)^{2}\right\rangle -\left(\langle a^{\dagger}a\rangle+\frac{1}{2}\right),\label{eq:13}
\end{equation}
where $Y_{\vartheta}=\frac{1}{2}\left(a^{\dagger2}e^{i\vartheta}+a^{2}e^{-i\vartheta}\right)$.
Considering the minimum of $Y_{\vartheta}$ with respect to phase
$\vartheta$, we have \citep{33,Mishra_2021}

\begin{align}
Y_{min} & =\langle a^{\dagger2}a^{2}\rangle-\vert\langle a^{2}\rangle\vert^{2}-\vert\langle a^{4}\rangle-\langle a^{2}\rangle^{2}\vert,\label{eq:14}
\end{align}
and AS squeezing occurs if $Y_{min}<0$. In order to compare AS squeezing
for states with different energy, we consider the renormalized factor 

\begin{equation}
S_{ass}=\frac{\frac{1}{2}\left[\langle a^{\dagger2}a^{2}\rangle-\vert\langle a^{2}\rangle\vert^{2}-\vert\langle a^{4}\rangle-\langle a^{2}\rangle^{2}\vert\right]}{\langle a^{\dagger}a+\frac{1}{2}\rangle}.\label{eq:16-1}
\end{equation}
AS squeezing occurs when $-1\le S_{ass}<0$.

\begin{figure}
\includegraphics[width=8cm]{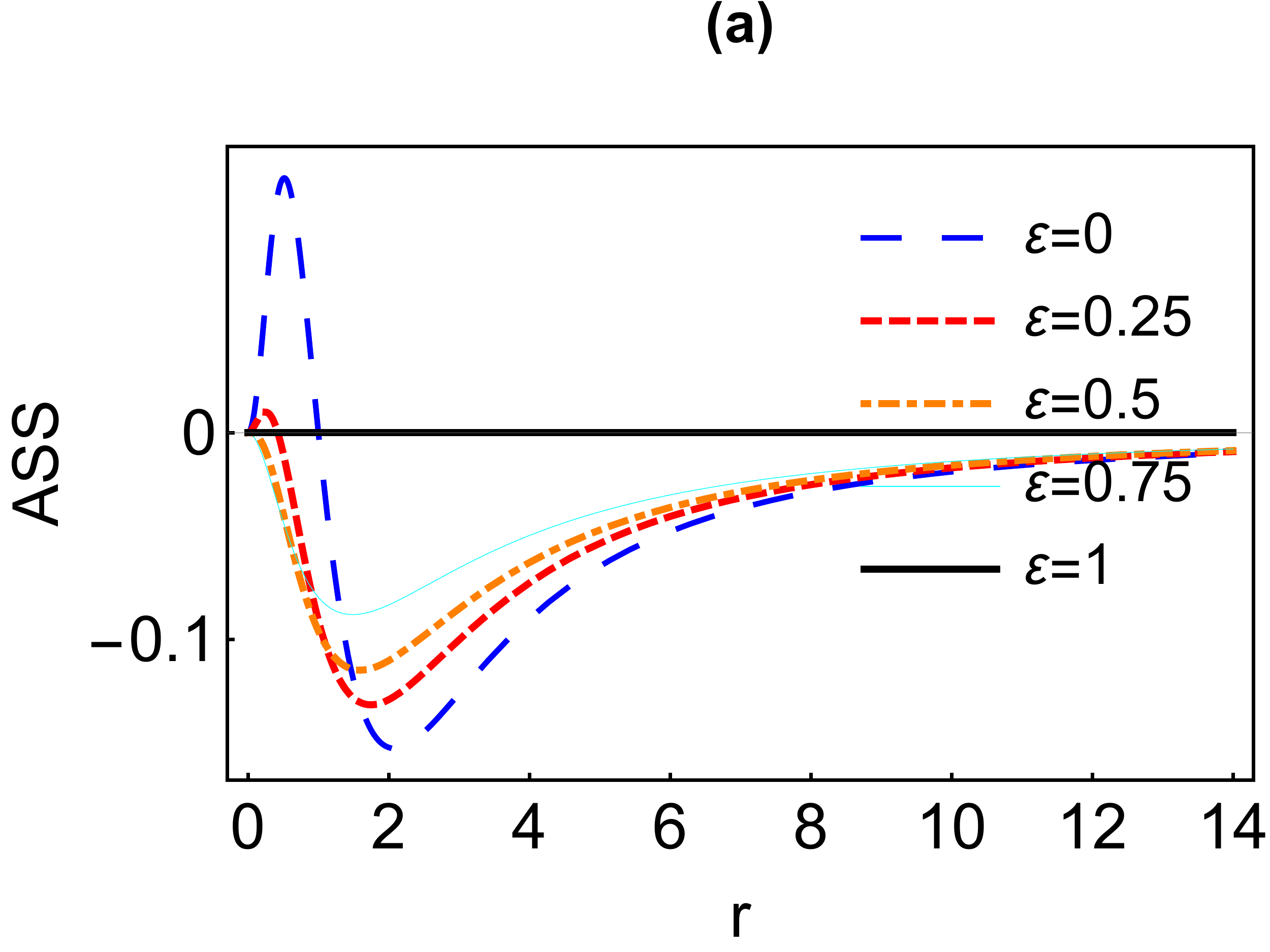}

\includegraphics[width=8cm]{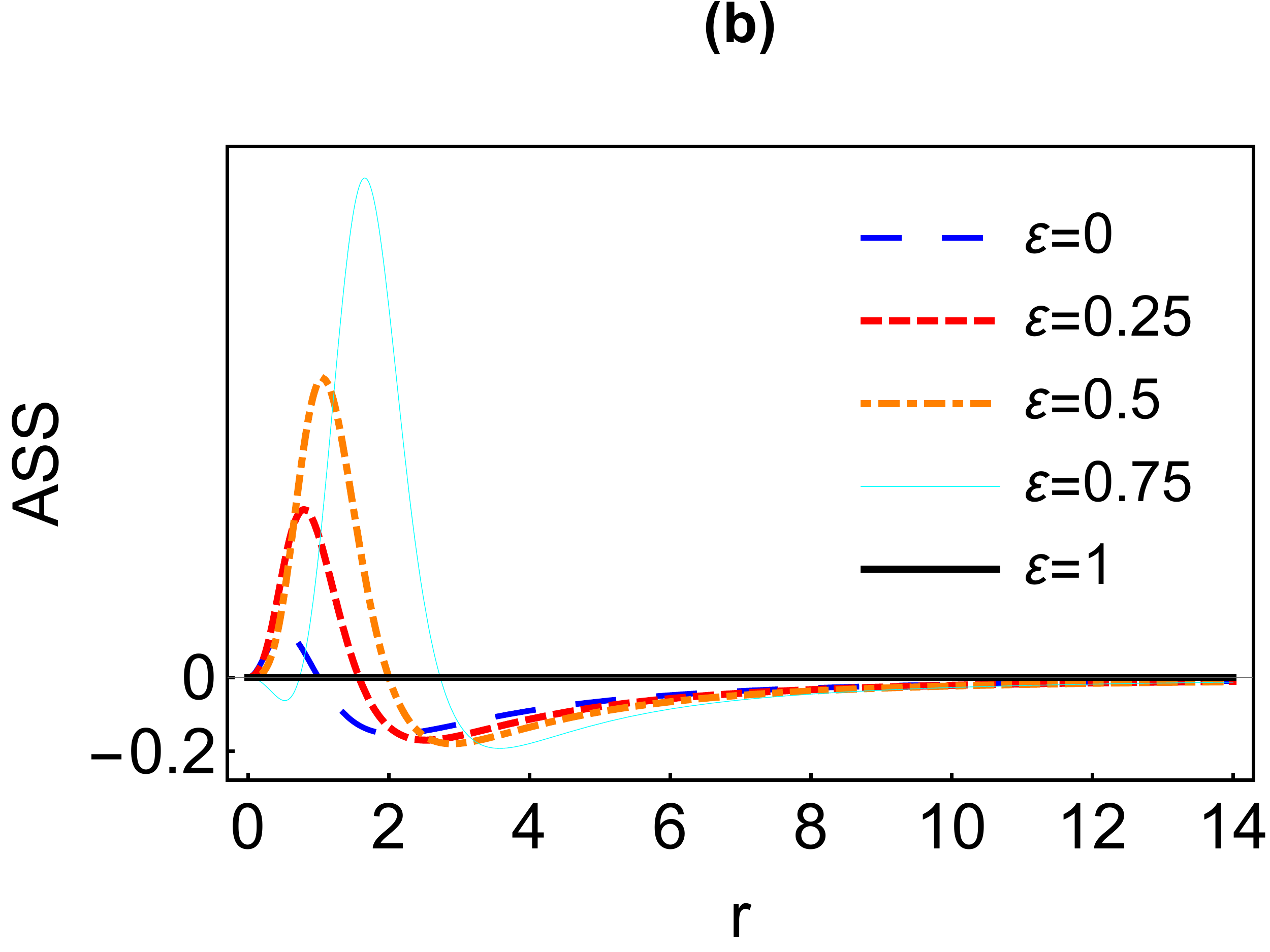}

\caption{\label{fig:4}AS squeezing parameter $S_{ass}$ of HCS as function
of coherent state parameter $\vert\alpha\vert$ for different $\varepsilon$.
Here, we take $\omega=\theta=0$, and $\varphi=0$ in (a), $\varphi=\pi$
in (b).}
\end{figure}

For the state $\vert\psi\rangle$, we calculate the explicit expressions
of the required quantities $S_{ass}$ . From Fig. \ref{fig:4} (a)
and (b), and by using the numerical plots of Eq. (\ref{eq:16-1}),
we observe the occurrence of ASS for different conditions. In Fig.
\ref{fig:4}(a), we plot the $S_{ass}$ as a function of $\vert\alpha\vert$
for different values of $\varepsilon$ while the remaining parameters
are fixed to $\omega=\varphi=0$ and $\theta=0$. The latter figure
highlights that the AS squeezing is significant as we increase $\vert\alpha\vert$
and has the same trend when $\vert\alpha\vert$ is even larger. From
Fig. \ref{fig:4}(a) we also observe that the AS squeezing of the
coherent superposition and the SPAC state always exist compared to
the SPAC state ($\varepsilon=0$ case in Eq. (\ref{eq:1})). Additionally,
in Fig. \ref{fig:4}(b), we plot the $S_{ass}$ with fixed $\varphi=\pi$,
and found that the effect of phase changing of the state defined in
Eq. ( \ref{eq:1}) on the AS squeezing is very obvious. It is evident
that for the $\varphi=\pi$ case, the AS squeezing diminished gradually
for $\varphi=0$, the $\vert\alpha\vert$ increased for all values
of $\varepsilon$ for except $\varepsilon=1$ (Fig. \ref{fig:4} (b)).
Furthermore, Fig. \ref{fig:4} (b) also indicated that the AS squeezing
of the state corresponding to the $\vert\alpha\vert>3$ region is
higher than the SPAC state.

We have to mention that assessing non-classicality of a given state
using negativity $Q_{m}$ and $S_{\phi}$ provides only sufficient
conditions \citep{Agarwal2013}. In other words, a non-classical field
may have $Q_{m}$ and $S_{\phi}$ both negative, only one negative,
or even both positive.

\subsection{Wigner function}

Next we study the non-classical features of HCS considering the negativity
of the Winger function, which is a quasi-probability distribution
function over the phase space. Its negativity is direct proof of the
state's non-classicality. To further confirm the non-classical features
of HCS, we investigate their Wigner function, which is defined as
the two-dimensional Fourier transform of the symmetrically ordered
characteristic function, i.e. \citep{Agarwal2013} 
\begin{equation}
W(z)\equiv\frac{1}{\pi^{2}}\int_{-\infty}^{+\infty}\exp(\lambda^{\ast}z-\lambda z^{\ast})C_{N}(\lambda)e^{-\frac{\lambda^{2}}{2}}d^{2}\lambda,\label{eq:35-1}
\end{equation}
where $C_{N}(\lambda)$ is the normally ordered characteristic function,
defined as: 
\begin{equation}
C_{N}(\lambda)=Tr\left[\rho e^{\lambda a^{\dagger}}e^{-\lambda a}\right].\label{eq:34}
\end{equation}
The explicit expression of the HCS Wigner function is 
\begin{align}
W(z) & =\frac{2\mathcal{N}^{2}}{\pi}e^{-2\vert\alpha-z\vert^{2}}\{\varepsilon-2\sqrt{\varepsilon-\varepsilon^{2}}w_{1}\label{eq:e6}\\
 & -(1-\varepsilon)(1-\vert2z-\alpha\vert^{2})\}\nonumber 
\end{align}
with 
\begin{equation}
w_{1}=Re\left[e^{i(\theta-\varphi)}(Re[\alpha]-2Re[z]+iIm[\alpha]-2iIm[z])\right].\label{eq:46}
\end{equation}
Here, $z=x+ip$ is a complex number in phase space. In general, the
Wigner function is real and bounded by $\vert W(z)\vert\le\frac{2}{\pi}$.
It can be easily check that if we take $\varepsilon=0$, the above
Wigner function is reduced to the Wigner function of SPAC state, i.e.,
\begin{equation}
W_{spacs}(z)=\frac{-2\left(1-\vert2z-\alpha\vert^{2}\right)}{\pi(1+\vert\alpha\vert^{2})}e^{-2\vert z-\alpha\vert^{2}}.\label{eq:234}
\end{equation}
On the other hand, for $\varepsilon=1$ Eq. (\ref{eq:35-1}) is reduced
to: 
\begin{equation}
W_{coh}(z)=\frac{2}{\pi}e^{-2\vert z-\alpha\vert^{2}},\label{eq:coh}
\end{equation}
which is the Wigner function of a coherent state. It can be seen that
the term with $w_{1}$ in Eq.(\ref{eq:e6}) is caused by the interference
between the coherent and SPAC components, and this term is our main
concern since it may lead to enhanced non-classical features in phase
space.

\begin{widetext} 

\begin{figure}

\includegraphics[width=15cm]{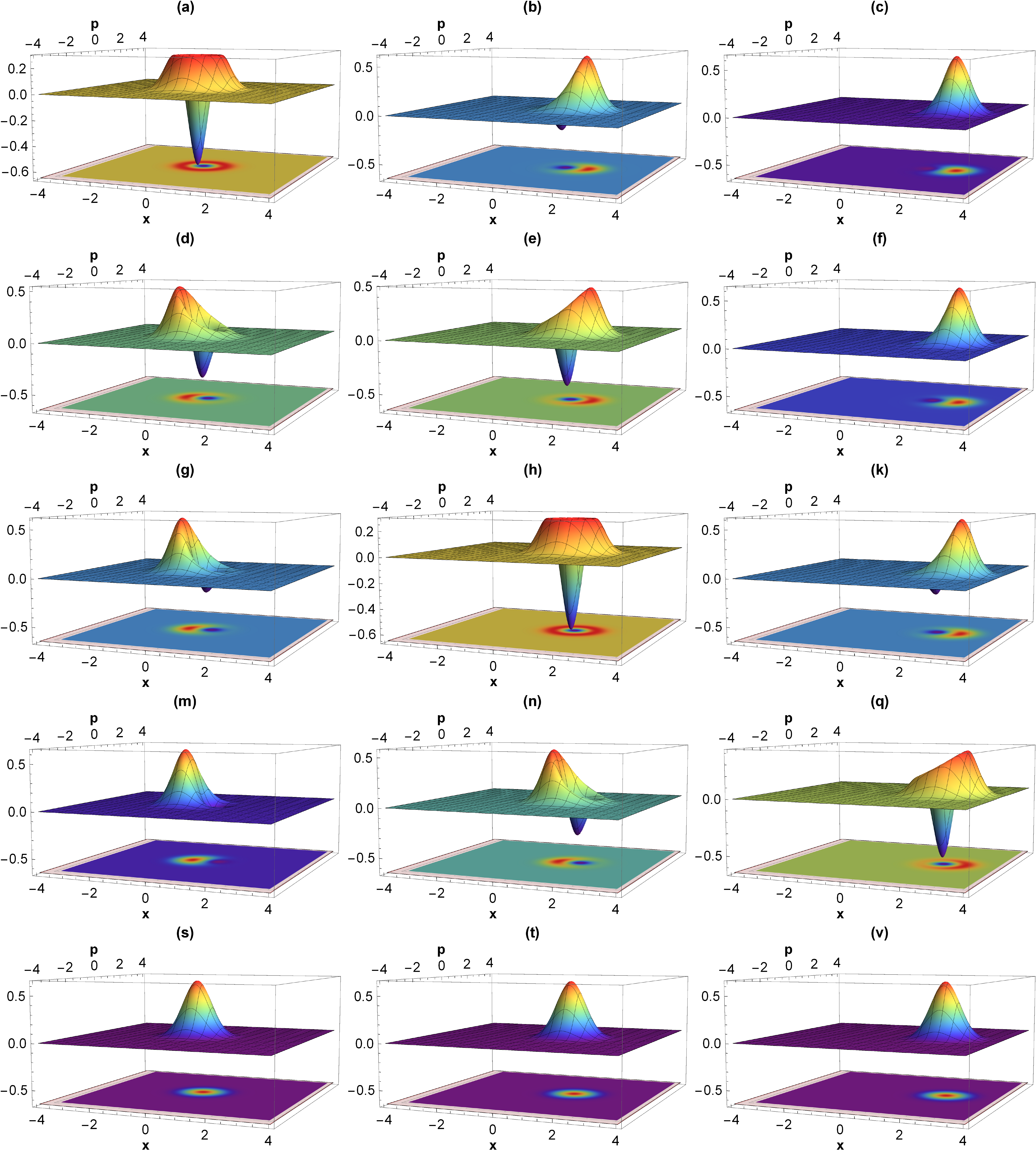}

\caption{\label{fig:5}Wigner function of different HCS states. The columns
show the result for $\vert\alpha\vert=0,1,2$), from left to right.
Panels (a) to (c) correspond to $\varepsilon=0$, (d) to (f) to $\varepsilon=0.25$,
(g) to (k) to $\varepsilon=0.5$, (m) to (q) to $\varepsilon=0.75$,
and (s) to (v) to $\varepsilon=1$, respectively. The other parameters
are set as in Fig.\ref{fig:1}.}

\end{figure}

\end{widetext}

The Wigner functions under various HCS are depicted in Fig. \ref{fig:5},
where the first and last rows, i.e., Fig.\ref{fig:5}(a) to (c) and
Fig.\ref{fig:5}(s) to (v) present the Wigner functions of SPAC and
the coherent states for the $\vert\alpha\vert=0,1,2$ cases, respectively.
The effect of the interference term in Eq. (\ref{eq:e6}) can be seen
in the plots of the second to fourth rows of Fig.\ref{fig:5}. The
negativity of the SPAC states decreases with the amplitude of $\vert\alpha\vert$
since the coherent approach states for large $\vert\alpha\vert$.
Note that the Wigner function of the coherent state is Gaussian {[}see
Fig. \ref{fig:5} (s) to (v){]}. However, as illustrated in Fig. \ref{fig:5},
the Wigner functions of a hybrid state, as defined in Eq. (\ref{eq:1})
for $\varepsilon\neq0,1$ present very good negativity. Specifically,
the negativity of HCS may be large, as illustrated in Fig. \ref{fig:5}
(h) for $\varepsilon=0.5$, and the state $\frac{1}{\sqrt{2}}\left(a^{\dagger}-1\right)\text{\ensuremath{\vert\alpha\rangle}}$
for $\vert\alpha\vert=1$ has same the Winger function as a single
photon (Fig. \ref{fig:5} (n)). Another interesting phenomenon is
that for $\varepsilon=0.75$ and large $\vert\alpha\vert$ values,
the peaks are further compressed along the $x$ direction at the expense
of an increase along the $p$ direction and can also exhibit a large
negative region. This indicates that the negativity of the corresponding
state increases as the coherent state amplitude $\vert\alpha\vert$
increases {[}see Fig. \ref{fig:5} (m) to (q){]}. This feature is
not possessed in either the SPAC or the coherent state. From the above
numerical analysis, we conclude that the superposition of the SPAC
state with a coherent state has a very good negativity of the Wigner
function for some parameters compared to the solely SPAC state. This
phenomenon infers that our states of $\vert\psi\rangle$, especially
for $\varepsilon=0.5,0.75$ has a strong non-classicalities compared
with the SPAC state.

\section{\label{sec:3}Possible realiztion of HCS}

\begin{figure}
\includegraphics[width=8cm]{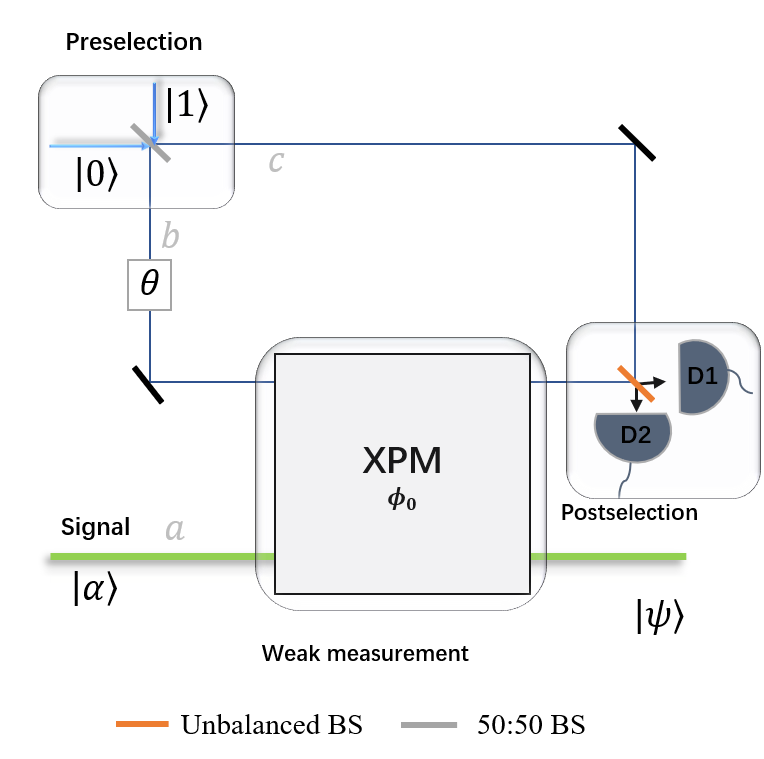}

\caption{\label{fig:6} Preparation scheme for HCS. The signal mode is prepared
in a coherent state $\vert\alpha\rangle$, and interact with a single
photon through cross Kerr medium. After the photon is detected at
$D1$ with nonzero probability, the signal mode is prepared in a superposition
of SPAC and coherent states. The same setup has been also suggested
to prepare phase-squeezed optical pulses \citep{RN2101} and to amplify
single-photon non-linearity \citep{PhysRevLett.107.133603} .}
\end{figure}

As mentioned in the Introduction, there are many feasible schemes
to prepare HCS. Here, we suggest another possible preparation scheme,
which is schematically illustrated in Fig. \ref{fig:6} and involves
a single photon coupled by the cross-Kerr effect to a signal mode.
We assume that the signal mode is prepared in a coherent state $\vert\alpha\rangle$,
and the single photon (probe mode) interacts with the signal through
a Kerr medium, leading to a cross-phase shift characterized by the
Hamiltonian: 
\begin{equation}
H_{I}=\hbar\chi a^{\dagger}ab^{\dagger}b=\hbar\chi\hat{n}_{a}\hat{n}_{b}.
\end{equation}
Here, $\hat{n}_{a}=a^{\dagger}a$ and $\hat{n}_{b}=b^{\dagger}b$
denote photon number operators in arms $a$ and $b$, respectively,
and $\chi$ is related to the third-order nonlinear susceptibility
$\chi^{(3)}$ of the medium. If the single photon is sent through
a 50-50 beam splitter, then the state of modes $b$ and $c$ is given
by 
\begin{equation}
\vert\psi_{i}\rangle=\frac{1}{\sqrt{2}}\left(\vert1\rangle_{b}\vert0\rangle_{c}+i\vert0\rangle_{b}\vert1\rangle_{c}\right).\label{eq:454}
\end{equation}
We add a phase shifter in mode \textbf{$b$} so that the state $\vert\psi_{i}\rangle$
before interacting with signal is changed to 
\begin{equation}
\vert\psi_{i}^{\prime}\rangle=\frac{1}{\sqrt{2}}\left(e^{i\theta}\vert1\rangle_{b}\vert0\rangle_{c}+i\vert0\rangle_{b}\vert1\rangle_{c}\right).\label{eq:35}
\end{equation}
After the Kerr interaction, the entire state of the system reads as
follows
\begin{align}
\vert\Psi^{\prime}\rangle & =U\vert\psi_{i}\rangle\otimes\vert\alpha\rangle=\exp\left(-i\phi_{0}\hat{n}_{a}\hat{n}_{b}\right)\vert\psi_{i}^{\prime}\rangle\otimes\vert\alpha\rangle\nonumber \\
 & \approx\left(1-i\phi_{0}\hat{n}_{a}\otimes\hat{n}_{b}\right)\vert\psi_{i}^{\prime}\rangle\otimes\vert\alpha\rangle.\label{eq:No post-1}
\end{align}
where $\phi_{0}=\chi t$ is the phase shift caused by the weak Kerr
cross-phase-modulation (XPM), which characterizes the coupling strength
between the measured system and the measuring device, $t$ is the
interaction time $t=L/\upsilon$, $L$ is the length of the medium
and $\upsilon$ is the velocity of light in the medium. In general,
$\phi_{0}$ is very small, and we may expand the above unitary to
the first order.

Then, the single photon in the two arms passes through a variable
beam splitter (VBS) with transmissivity $t$ and reflectivity $r$
with $t^{2}+r^{2}=1$. If the photon is detected by the detector $D1$
at the output port, the quantum state of the single photon is post-selected
to be in \citep{RN2101} 

\begin{align}
|\psi_{f}\rangle_{1d} & =t\vert1\rangle_{b}\vert0\rangle_{c}+r\vert0\rangle_{b}\vert1\rangle_{c}.\label{eq:56-1}
\end{align}
 When the single photon is detected at the photon detector $D1$,
the entire state $\vert\Psi^{\prime}\rangle$ is post-selected to
$\vert\psi_{f}\rangle_{1d}$ and the final state of the signal mode
changes to
\begin{align}
\vert\Phi\rangle & =\langle\psi_{f}\vert\Psi^{\prime}\rangle\nonumber \\
 & =\frac{1}{\sqrt{2}}\left(e^{i\theta}t+ir\right)\vert\alpha\rangle-ie^{i\theta}\frac{t\phi_{0}\alpha}{\sqrt{2}}a^{\dagger}\vert\alpha\rangle\label{eq:44}
\end{align}
with 
\begin{align}
\langle\hat{n}_{b}\rangle_{\psi_{f},\psi_{i}} & =\langle\psi_{f}\vert b^{\dagger}b\vert\psi_{i}\rangle=\frac{te^{i\theta}}{\sqrt{2}}.\label{eq:45}
\end{align}
If we modify the phase shifter such that $\theta=-\frac{\pi}{2}$,
then Eq. (\ref{eq:44}) can be rewritten as (unnormalized) 
\begin{align}
\vert\psi\rangle & =i\frac{r-t}{\sqrt{2}}\vert\alpha\rangle+\frac{t\phi_{0}\alpha}{\sqrt{2}}a^{\dagger}\vert\alpha\rangle.\label{eq:31}
\end{align}
It is evident that Eq.(\ref{eq:31}) has the same form as Eq. (\ref{eq:1}),
and one can accomplish the state transition from a coherent state
to SPAC state by adjusting the transmissivity $t$ of VBS.

\section{\label{sec:4}Conclusion remarks}

This paper studies the non-classical properties of coherent hybrid
states by evaluating and assessing different non-classical quantifiers
such as squeezing, sub-Poissonianity, and negativity of the Wigner
function. We have found that the state $\vert\psi\rangle$ for $\varepsilon\neq0,1$
have relevant non-classical properties, sometimes large than the corresponding
SPAC state. Especially for $\varepsilon=0.5,0.75$, the corresponding
states show a large amplitude and AS squeezing as well as negative
regions of Wigner functions and those features still exist for larger
$\vert\alpha\vert$. The non-classical properties may be tuned by
varying the superposition parameter $\varepsilon$, which depends
on the amplifier strength \citep{PhysRevA.100.043802} and the beam-splitter
transmittance \citep{PhysRevA.82.053812,PhysRevA.91.022317,PhysRevA.98.013809,PhysRevA.104.033715,PhysRevA.105.022608}. 

The SPAC states have different applications, which are, however, restricted
to small amplitude (since, for large amplitude, they approach coherent
states). In this framework, our results show that non-classicality
and non-Gaussianity may be preserved at larger amplitudes by superimposing
the SPAC to coherent states.

On the other hand, with the rapid development of techniques for making
higher-order correlation measurements in quantum optics and laser
physics, it is possible to define the higher-order squeezing effect
of the field. We have also observed that AS (second order) squeezing
of HCS for $\varepsilon\neq0,1$ cases are enhanced compared with
SPAC state. Thus, we anticipate that this hybrid coherent state could
be used in quantum information processing and may help improve associated
schemes' performance.
\begin{acknowledgments}
This work was supported by the National Natural Science Foundation
of China (Grants No. 11865017).
\end{acknowledgments}

\bibliographystyle{apsrev4-1}
\bibliography{ref}

\end{document}